%% file: main-NP.tex
\documentclass[twocolumn, showpacs, english, preprintnumbers, amsmath, amssymb, superscriptaddress, aps, prl,longbibliography,nofootinbib,10pt]{revtex4-2}
\usepackage{silence}
\usepackage{xspace}

\usepackage[utf8]{inputenc}
\usepackage{graphicx}
\usepackage{grffile}
\usepackage{amssymb}
\usepackage[dvipsnames]{xcolor}
\usepackage{amsmath,bm}
\usepackage[normalem]{ulem}
\usepackage{appendix} 
\definecolor{orange}{rgb}{1,0.5,0}
\definecolor{goodgreen}{rgb}{0.1,0.5,0}
\definecolor{goodred}{rgb}{0.7,0,0}

\renewcommand\vec{\boldsymbol}
\usepackage{comment}

\usepackage{tikz}
\makeatletter

\usepackage[colorlinks,urlcolor=goodgreen,citecolor=blue,linkcolor=goodred]{hyperref}

\newcommand{\orcid}[1]{\href{https://orcid.org/#1}{\includegraphics[width=8pt]{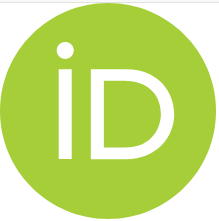}}}

\usepackage{color}
\newcommand{\Tau}{\mathrm{T}}

\definecolor{TTH-color}{rgb}{0.0,0.0,1}

\definecolor{SB-color}{rgb}{1.0,0.0,0}

\definecolor{RO-color}{rgb}{0.5,0.5,0}

\definecolor{TK-color}{rgb}{0.3,0.05,0.4}

\usepackage{float}

\let\oldepsilon\epsilon \let\epsilon\varepsilon \let\varepsilon\oldepsilon

\newcommand\commcircsym{\raisebox{-.5ex}{\shortstack{%
  \(\circ\)\\[-.5ex]%
  \(,\)}}%
}
\newcommand\commcirc{\protect\commcircsym}
\makeatother

\newcommand{\vdw}{v_{\mathrm{dw}}}
\newcommand{\qex}{q_{\mathrm{ex}}}
\newcommand{\dFI}{d_{\mathrm{FI}}}
\newcommand{\dSC}{d_{\mathrm{SC}}}
\newcommand{\IW}{I_{\mathrm{W}}}
\newcommand{\AF}{A_{\mathrm{F}}}
\newcommand{\AS}{A}

\usepackage{babel}
\makeatother

\begin{document}

\title{Controlling magnetic domain walls with supercurrents}
\author{Tim Kokkeler  \orcid{0000-0001-8681-3376}} 
\email{tim.h.kokkeler@jyu.fi}
\affiliation{These authors contributed equally to the work.}
\affiliation{Department of Physics and Nanoscience Center, University of Jyväskylä, P.O. Box 35 (YFL), FI-40014 University of Jyväskylä, Finland}

\author{Risto Ojajärvi 
 \orcid{0000-0001-9665-6503}}
 \email{risto.m.m.ojajarvi@jyu.fi}
 \affiliation{These authors contributed equally to the work.}
\affiliation{Department of Physics and Nanoscience Center, University of Jyväskylä, P.O. Box 35 (YFL), FI-40014 University of Jyväskylä, Finland}

\author{F. Sebastian Bergeret\orcid{0000-0001-6007-4878}}
\email{fs.bergeret@csic.es}
\affiliation{Centro de Física de Materiales (CFM-MPC) Centro Mixto CSIC-UPV/EHU,
E-20018 Donostia-San Sebastián, Spain}
\affiliation{Donostia International Physics Center (DIPC), 20018 Donostia--San
Sebastián, Spain}

\author{Tero T. Heikkilä \orcid{0000-0002-7732-691X}}
\email{tero.t.heikkilä@jyu.fi}
\affiliation{Department of Physics and Nanoscience Center, University of Jyväskylä, P.O. Box 35 (YFL), FI-40014 University of Jyväskylä, Finland}

\begin{center}

\end{center}

\begin{abstract}
    Establishing a versatile, fast and reliable magnetic memory technology is a giant bottleneck for cryogenic computing since present-day room-temperature solutions either cease to work or consume too much power. The long-term goal of superconducting spintronics has been to overcome this bottleneck by generating magnetic memories with equal-spin triplet supercurrent driven through them to control their magnetization direction. This path has been hampered by the short spin relaxation length and strong anisotropy in ferromagnets. Here we show how the supercurrent driven  generation of spin accumulation in a superconductor/magnetic insulator bilayer, together with Gilbert damping of magnetization lead to a motion of magnetic domain walls. 
    This manifests as a local voltage across the wall, which allows its position to be identified. Associated with this voltage and the current, there is Joule power which is dissipated via the Gilbert damping. The power required to maintain domain wall motion is orders of magnitude smaller than 
    in the normal state,
    where most of the power is wasted in producing the current. 
\end{abstract}
\maketitle
\begin{figure}
    \centering
    \includegraphics[width=1\linewidth]{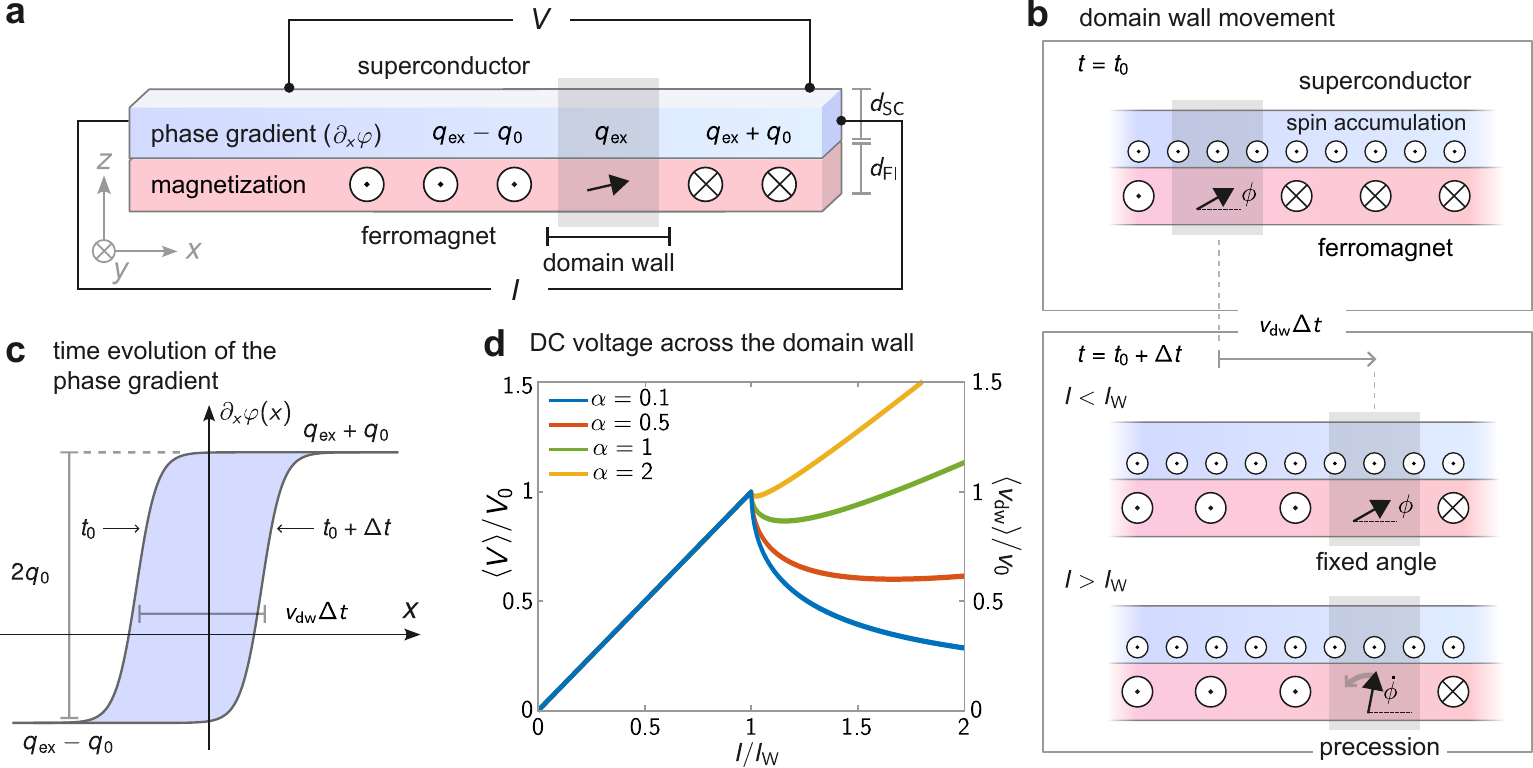}
    \caption{(\textbf{a}) Superconductor--ferromagnet bilayer. Magnetic proximity effect and SOC induce into the superconductor a helical superfluid momentum $q_0$, whose sign changes across the domain wall. Supercurrent $I$ induces an additional superfluid momentum $q_{\mathrm{ex}}$. (\textbf{b}) Spin accumulation generated by the supercurrent via the spin-galvanic effect drives the domain wall (DW) at velocity $v_{\mathrm{dw}}$. Below the Walker breakdown current, $I<\IW$, the wall moves with a fixed shape, while above the Walker breakdown current, $I>\IW$, the spins in the domain wall precess while the wall moves. (\textbf{c}) Snapshots of the superconducting superfluid momentum at two distinct times during DW motion. The voltage $V = 2q_0 v_{\mathrm{dw}}$ across the DW is proportional to the area between the curves. (\textbf{d}) DC voltage/average DW velocity as a function of the driving current for different values of Gilbert damping $\alpha$. Here, $v_0 = \lambda K_\perp/4\hbar$, $V_0= 2 \frac{\hbar}{e}q_0 v_0$, and $\IW \propto \alpha$, with a constant of proportionality defined in Eq.~\eqref{eq:Walker_current}.}
    \label{fig:racetrack}
\end{figure}

The prediction of equal-spin triplet superconducting correlations \cite{bergeret2001long} in ferromagnets, and their subsequent experimental confirmation through long-range Josephson effects in magnetic junctions \cite{khaire2010observation, robinson2010controlled, sprungmann2010evidence
}, 
motivated the introduction of the concept of superconducting spintronics \cite{birge2018spin, 
linder2015superconducting, yang2021boosting}. In analogy with conventional spintronics, spin-polarized supercurrents were expected to provide efficient torques for controlling magnetic textures. However, despite extensive experimental efforts \cite{jeon2020tunable,martinez2020interfacial,jeon2021role,bhatia2021nanoscale,jeon2023chiral,chan2023controlling,bregazzi2024enhanced},  
the practical realization of superconducting spin-torque functionalities has remained largely elusive.
 This is mainly because generating 
 equal-spin 
 supercurrents 
 typically requires complex non-collinear magnetic multilayer structures, together with long spin-diffusion lengths and high current densities to produce appreciable torques \cite{yang2021boosting}. 

Here, we propose a novel 
setup that overcomes these difficulties and enables the control of magnetic textures by supercurrents without the need for equal-spin supercurrents. The system consists of a superconductor with spin–orbit coupling 
in contact with
a ferromagnetic insulator hosting a domain wall (Fig.~\ref{fig:racetrack}). An electric current $I$ is driven solely through the superconductor, thereby avoiding the limitations associated with magnetic Josephson-junction-based setups.

In this configuration, 
the supercurrent is converted into an equilibrium spin accumulation at the ferromagnetic-insulator interface via the inverse spin-galvanic effect \cite{mineev1994helical,edelstein1995magnetoelectric,konschelle2015theory}. The induced spin interacts with the ferromagnetic insulator via interfacial exchange coupling, exerting a torque that drives the magnetic domain wall.


In the superconductor, the motion of the domain wall generates a voltage via the spin-galvanic effect, which is enabled by the combination of spin-orbit coupling and the magnetic proximity effect. In the static configuration, these ingredients give rise to helical superconductivity, characterized by a spontaneous superfluid momentum $\hbar q_0$ in the 
zero-current ground state
\cite{ganichev2019spin}. Since $q_0$ is odd under magnetization reversal, 
it has opposite signs on the two sides of the domain wall.
In contrast, an externally driven supercurrent is associated with a uniform superfluid momentum
$\qex = \frac{\hbar}{e} I /(D_S \AS)$, where $D_S$ is the superfluid weight in the superconductor with cross-section $\AS$. As the domain wall moves, the superfluid momentum changes by $2\hbar q_0$ at positions where the magnetization reverses direction. This translates into a global time-dependent phase difference across the domain wall. Via the ac Josephson relation, this manifests as a measurable voltage $V=\frac{\hbar}{e} q_0 \vdw$, where $\vdw$ is the domain wall velocity, and a resistance $R \propto q_0^2$. 
This means that supercurrent-driven domain wall motion is a nonequilibrium effect which requires power $P=I V$ to maintain the current. This power is dissipated through damping in the ferromagnetic insulator.
As the resistance is only due to domain wall motion, the heating is orders of magnitude weaker than in corresponding normal-metal setups.

Strictly speaking, the microscopic modeling of domain wall motion requires a nonequilibrium theory. Such a description, based on the Keldysh-Usadel theory, is provided in the Methods section. In the main text we focus on the adiabatic limit.

\section*{Force on the domain wall from the inverse spin galvanic effect}
 \label{sec:dwforce}
 
We consider a wire constructed from a bilayer of a superconductor (SC) and a ferromagnetic insulator (FI), as shown in Fig.~\ref{fig:racetrack}(a). We assume that the wire is much longer than other length scales, such as the 
domain wall core size $\lambda$, spin relaxation length $\ell_{\rm so}$, superconducting coherence length $\xi_0$ and charge relaxation length $\ell_{c}$ in the superconductor. Moreover, we assume that the system is homogeneous on either side of the wall, and that the thickness of the bilayer is small enough that spin effects dominate over orbital effects related to the Oersted field, see Methods.

The ferromagnet is described by the free energy
 \cite{abert2019micromagnetics}
\begin{align}\label{eq:FFI}
    F_{\text{FI}} = \nu_s\int dV  J(\partial_x m)^2-K_{\parallel}m_y^2+K_{\perp}m_z^2\;,
\end{align}
where $\vec{m}=(m_x,m_y,m_z)$ is the unit vector along the direction of the magnetization, $\nu_s$ is the density of spins, $K_{\parallel,\perp}$ are the anisotropy energies of the easy and hard axes and $J$ is the spin stiffness. To maximize the efficiency of the spin galvanic effect, we choose the easy axis perpendicular to the supercurrent and the domain wall direction along with it. The physics is analogous for N\'eel and Bloch walls.

We assume the superconductor is a conventional $s$-wave superconductor with spin-orbit coupling (SOC). Microscopically, the SOC can be either intrinsic,originating from inversion symmetry breaking, 
or extrinsic, generated by scattering from impurities with SOC \cite{bergeret2016manifestation,virtanen2021magnetoelectric}. 
We describe a diffusive superconductor 
exploiting the Usadel formalism and its free energy \cite{virtanen2020quasiclassical},  applicable from low temperatures to above the critical temperature $T_c$ and in a non-equilibrium setting. We do not expect the results to change qualitatively towards the clean limit.

The FI/SC interface is described by the energy contribution
\begin{align}\label{eq:interfacefreeenergy}
    F_{\text{SC-FI}} = -\frac{G_i}{e^2\nu_0} \vec{m}\cdot\boldsymbol{S}\,,
\end{align}
where $G_i$ is the imaginary part of the spin-mixing conductance of the interface \cite{brataas2000finite}, $\nu_0$ is the normal-state density of states per spin in the superconductor, and $\vec{S}$ is the excess spin in the superconductor at the interface. In the superconductor, this coupling generates an effective exchange field \cite{meservey1994spin,bergeret2018colloquium,heikkila2019thermal,hijano2021coexistence}. The combination of time-reversal symmetry breaking (from the FI), inversion symmetry breaking (from the FI/SC interface), and spin-orbit coupling leads to an equilibrium superfluid momentum
$\hbar \vec{q}_0$, where
\begin{align}
    \vec{q}_0 = \frac{G_i}{\sigma} \vec{m}\times \kappa\vec{n}\;,\label{eq:equilibriumphase}
\end{align}
$\vec{n}$ is the normal to the FI/SC interface, outward from the superconductor, and $\sigma$ is the normal-state conductivity of the superconductor. The parameter $\kappa$ is a dimensionless parameter that depends on the setup and spin-orbit coupling strength. This general phenomenology gives rise to a spin-galvanic effect with either intrinsic or extrinsic SOC. The equilibrium superfluid momentum is odd in the magnetization and therefore has opposite sign on different sides of the wall (Fig.~\ref{fig:racetrack}a). Its magnitude is maximized if the spins in the domains of the ferromagnet are perpendicular to both the interface normal and the transport direction.

For intrinsic SOC, such as Rashba SOC, if the spin-orbit relaxation length $l_{\text{so}}$ of the superconductor is much shorter than the superconductor coherence length, the parameter $\kappa$ is given by $\kappa = \frac{\gamma l_{\text{so}}^2}{Dd_{\text{SC}}}$, where  $\gamma$ is the spin-galvanic coefficient \cite{ganichev2019spin,kokkeler2025universal}, $D$ is the diffusion constant in the superconductor and $l_{\text{so}}$ is the spin-relaxation length due to SOC in the superconductor. For materials with a spin-Hall angle  $\theta$, in the same limit, $\kappa = \theta \frac{l_{\text{so}}}{d_{\text{SC}}}\tanh{\frac{d_{\text{SC}}}{2l_{\text{so}}}}$ (see Methods).

\begin{figure}
    \centering
    \includegraphics[width=1\linewidth]{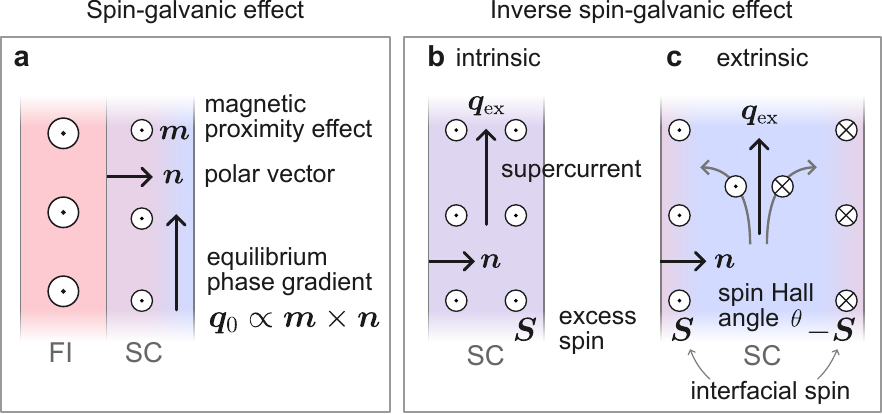}
    \caption{(\textbf{a}) Generation of equilibrium superfluid momentum by spin-galvanic effect. Magnetic proximity effect induces a Zeeman field in the SC (with direction $\vec{m}$). Together with the polar vector $\vec n$ this generates a current, which is canceled by a superfluid momentum $\vec{q}_0$. (\textbf{b}) and (\textbf{c}) Generation of spin by supercurrent due to inverse spin galvanic effect (ISGE). The intrinsic ISGE generates an excess spin in the bulk of the SC due to broken inversion symmetry. The extrinsic ISGE (spin Hall effect) generates a transverse spin current which shows up as interfacial spin. Both mechanisms involve the polar vector $\vec n$.}
    \label{fig:SGE}
\end{figure}

The existence of a spin-galvanic effect implies an inverse effect (Fig.~\ref{fig:SGE}b--c), the creation of spin by an externally applied supercurrent $\vec j = D_S \vec q_{\text{ex}}$ \cite{edelstein1995magnetoelectric, hals2016supercurrent}:
\begin{align}
    \vec{S} =  \chi_s \kappa \vec{n}\times\vec{j}\;,\label{eq:inverseSGE}
\end{align}
where $\chi_s$ is a local spin response coefficient, $D_S$ is the superfluid weight and $\vec{q}_{\text{ex}}$ is the deviation of superfluid momentum from its equilibrium value. As discussed in Methods, if the spin-relaxation length is shorter than the superconductor coherence length, we have $\chi_s = \hbar d_{\text{SC}}/e D$ at the interface. 
According to Eq.~\eqref{eq:inverseSGE}, the spin  is proportional to the applied current and independent of the ferromagnetic insulator, which means that the created spin is homogeneous along the transport direction (Fig.~\ref{fig:racetrack}b). It is an equilibrium spin resulting from ground state correlations, and does not consist of excited quasiparticles.

Following the FI/SC coupling, Eq.~\eqref{eq:interfacefreeenergy}, electrons  exert a torque on the spins in the wall,
\begin{align}\label{eq:Torque}
     \mathcal{T} = -\frac{G_i}{e^2\nu_0}\vec{m}\times\vec{S}\;,
\end{align}
while the spins exert exactly the opposite torque on the electrons. 
If the magnetization far away from the  wall is perpendicular to both $\vec n$ and the applied current $\vec j$, this torque is non-zero only at the wall, inducing a rotation of the domain wall spins on $xz$-plane. As shown in Methods, the torque on the electrons is exactly counterbalanced by a torque associated with the spin-galvanic effect, consistent with the absence of spin currents on either side of the wall. 

\section*{ Supercurrent-driven domain wall dynamics}
\label{sec:dwdynamics}

The full magnetization dynamics can be determined from the variation of an action corresponding to the above free energies, in combination with a dissipation functional to describe the interaction with an external bath, such as phonons. This gives the Landau-Lifshitz-Gilbert (LLG) equation \cite{gilbert2004phenomenological}. The torque, Eq.~(\ref{eq:Torque}), enters as a source term in the LLG equation. 

Solving the coupled Usadel and LLG equations provides the full dynamics of the system. However, further simplifications can be made assuming a strong easy-axis anisotropy in the ferromagnetic insulator. Under this assumption, the wall shape remains rigid during motion, allowing its dynamics to be described by two collective coordinates: the domain wall center position $x_0$ and the azimuthal angle $\phi$ of the magnetization in the $xz$-plane \cite{
thiele1973steady, tatara2008microscopic}. The equations of motion are:
\begin{align}
    -m_y^R\frac{1}{\lambda}\partial_t x_0 + C_1\alpha \partial_t \phi + C_1K_{\perp}\sin 2\phi + \frac{1}{\lambda}\tau = 0\label{eq:DWeq1}\;,\\
    m_y^R\partial_t \phi + C_2\frac{\alpha}{\lambda} \partial_t x_0 + f = 0\;.\label{eq:DWEq2}
\end{align}
Here $\lambda$ is the domain wall length, $\alpha$ is the external Gilbert damping parameter, $K_{\perp}$ is the hard axis anisotropy, $m_y^R= \pm 1$ indicates whether the magnetization on the right side of the wall is in the $\pm\hat{y}$-direction, while $C_1 = \frac{1}{\lambda}\int_{-\infty}^{\infty}dx(1-m_y^2)$ and $C_2 = \lambda\int_{-\infty}^{\infty}dx (\partial_x \vec{m})^2$ are dimensionless constants. For a standard domain wall shape \cite{tatara2008microscopic} they evaluate to $2$ and $\frac{12}{5}$ respectively.
The domain wall force $f$ and torque $\tau$ are determined from Eqs.~\eqref{eq:inverseSGE} and \eqref{eq:Torque} as (see Methods) 
\begin{align}
    \tau & = \frac{G_i}{e^2\nu_0\nu_s d_{\text{FI}}} \int dx\; \vec{S}\cdot(\vec{m}\times\vec{y}) = 0\;,\label{eq:tau}\\
    f &= \frac{G_i}{e^2\nu_0\nu_s d_{\text{FI}}} \int dx \;\vec{S}\cdot\partial_x\vec{m} = -\frac{d_{\text{SC}}}{d_{\text{FI}}}\frac{q_0 j}{e\nu_s}\;.\label{eq:force}
\end{align}
Here $d_\text{FI}$ is the thickness of the ferromagnetic insulator, while $q_0 = m_y^R G_i \kappa/\sigma$ is the equilibrium superfluid momentum far away from the wall on the right side.

Moreover, because the contribution to Gilbert damping from the electrons in the superconductor is exponentially suppressed at low temperatures, for domain wall speeds $\vdw \ll \lambda \Delta/\hbar$ we may ignore the excitation of quasiparticles in the superconductor. 
In this adiabatic limit, $\vec S$ is given by Eq.~\eqref{eq:inverseSGE}, and the domain wall velocity can be evaluated explicitly. Without Gilbert damping, the spin at the center of the wall precesses around $\vec S$, but there is no domain wall movement. Damping aligns this spin with the spin in the superconductor. Since it is parallel to the magnetization on the left (right) of the wall, this leads to domain wall movement towards the right (left) \cite{tatara2008microscopic}.

The resulting motion is characterized by two regimes separated by the Walker breakdown (Fig.~\ref{fig:racetrack}d) \cite{schryer1974motion} with a characteristic current
\begin{align}
I_{\mathrm{W}} = \frac{A C_1 C_2 \alpha e\nu_s K_{\perp} d_{\text{FI}}}{2q_0d_{\text{SC}}}\;.\label{eq:Walker_current}
\end{align} 
For currents smaller than $I_{\mathrm{W}}$, the wall moves with a constant velocity and the angle $\phi$ remains constant because the force of the electrons is counterbalanced by the hard-axis anisotropy. Above $\IW$, $\phi$ becomes time-dependent and the velocity oscillates. Below the threshold, the domain wall velocity can be expressed as 
\begin{align}
    \vdw = \frac{q_{0}\lambda}{C_2\alpha}\frac{d_{\text{SC}}}{d_{\text{FI}}}\frac{j}{e \nu_s}\;.\label{eq:velocity}
\end{align}
This equation is the central result of our paper. It shows that domain walls in ferromagnetic insulators are moved by supercurrents if the superconductor has an equilibrium superfluid momentum, Eq.~(\ref{eq:equilibriumphase}), through intrinsic spin-orbit coupling or a finite spin-Hall angle. The domain wall velocity increases with a better contact between the superconductor and the ferromagnetic insulator, and is limited by Gilbert damping and by spin-relaxation in the superconductor. Moreover, since it  is proportional to $q_0$, its sign is determined by the orientation of the magnetization on the right side of the wall. Thus, in systems with two walls, they either repel or attract, depending on the direction of the current and the magnetization between the walls.

The mechanism that leads to domain wall movement is based on the generation of equilibrium spin by the inverse spin-galvanic effect. This is an entirely different mechanism than those explored before \cite{takashima2017adiabatic,bobkova2018spin,rabinovich2019resistive}, which relied on spin-polarized supercurrents carried by equal-spin triplets. In our proposal they are not required. In fact, in the orientation studied here the spin $\vec{S}\propto \vec n \times \vec j$ generated in the superconductor is parallel to the easy axis, so that there are no equal-spin triplets far away from the domain wall.    If $\vec n$ and $\vec j$ were not perpendicular to the easy axis, $\vec S$ would acquire a perpendicular component, resulting in a nonzero domain wall torque $\tau$ in Eq.~\eqref{eq:tau}. This would have the same effect as applying a dc magnetic field perpendicular to the easy axis \cite{sobolev1995domain}. A perpendicular field cannot drive the domain wall, but modifies the dynamics by creating a preferred direction for the spin at the center of the wall.

\section*{Local electrical signatures of the domain wall motion}
\label{sec:localvoltage}
\begin{figure*}
    \centering
    \includegraphics[width=0.8\linewidth]{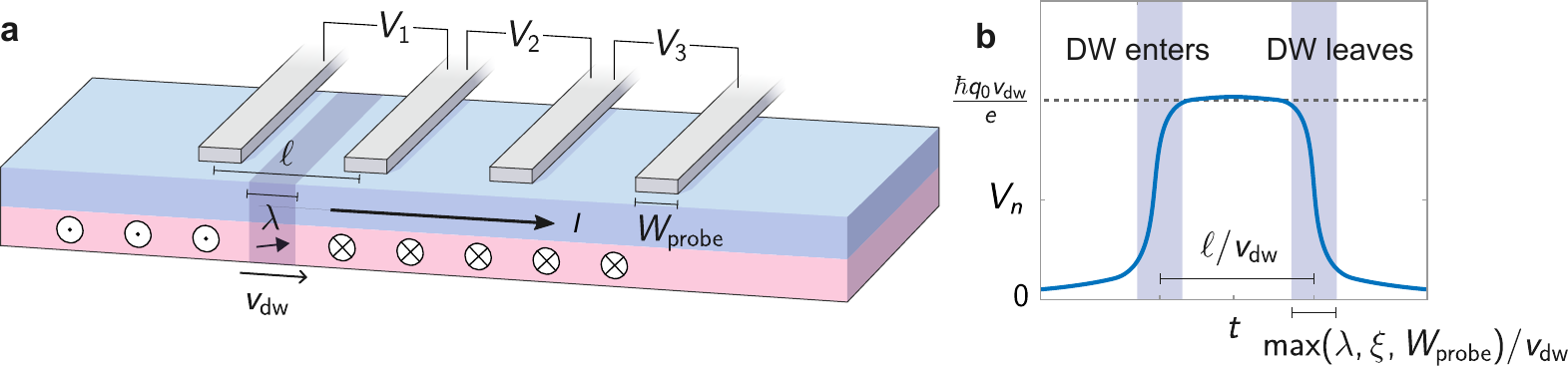}
    \caption{Electrical detection of the domain wall motion. (\textbf{a}) Setup with multiple voltage probes. (\textbf{b}) Sketch of the time-dependent voltage signal due to domain wall movement. The voltage is maximized when the domain wall is between the two probes. When the domain wall leaves, the signal decays sharply over a time-scale determined by the domain wall speed, and the domain wall size or the superconductor coherence length. After this initial period, a tail remains, which is exponentially suppressed in $\frac{\Delta}{k_B T}$, and extends over the charge imbalance relaxation time.}
    \label{fig:probe}
\end{figure*}
As illustrated in Fig.~\ref{fig:racetrack}c, domain wall movement leads to a time-dependent phase difference $\partial_t \varphi = 2q_0 \vdw$ across the wall. As in Josephson junctions, this implies a voltage drop over the wall:
\begin{align}
    V = \frac{\hbar}{e} q_0 \vdw\;.\label{eq:Voltagegeneration}
\end{align}
This voltage can be measured using the local probe configuration in Fig.~\ref{fig:probe}. Consider two of the probes and a domain wall moving from left to right, at temperatures well below the superconducting gap. If the wall is to the left of both probes, the current flow between the probes is  dissipationless and no voltage appears. When the  wall reaches the left probe, the voltage increases towards $\frac{\hbar}{2e}q_0 \vdw$ over a time-scale $\text{max}(\lambda, \xi_0, W_{\text{probe}})/\vdw$, where $\xi_0$ is the superconductor coherence length and $W_{\text{probe}}$ is the probe width. When the wall passes the right probe the voltage decays towards zero over the same time scale. 

At higher temperatures there are additional features in the voltage signal due to the generation of quasiparticles near the wall. These quasiparticles are partially relaxed over the superconductor coherence length or spin relaxation length, but due to the spin-orbit coupling, there is also a local quasiparticle current that decays over the charge imbalance length \cite{tokatly2025spin}. 
As voltage probes are sensitive to the quasiparticle potential, this effect leads to additional tails in the signal, 
as illustrated in Fig.~\ref{fig:probe}b.


\section*{Power dissipation and heat balance}
\label{sec:power}

The appearance of a voltage across the domain wall suggests that a power $P=IV$ is needed to sustain the current. This can be shown explicitly using heat balance equations, which result in 
(see Methods for details):
\begin{align}
    P = I V = \AS \frac{\hbar}{e^2}\frac{q_0^2 \lambda}{C_2 \alpha}\frac{\dSC}{\dFI}\frac{j^2}{\nu_s}\;.\label{eq:Power}
\end{align}
Below the Walker threshold, the dissipation rate 
in the ferromagnet is
\begin{align}
    D = \hbar \nu_s \AF \alpha \int dx (\partial_t \vec{m}) = \AF C_2\hbar \alpha\frac{\nu_s}{\lambda} \vdw^2\;,\label{eq:dissipation}
\end{align}
where $\AF$ is the cross section of the ferromagnet. Substituting Eq.~\eqref{eq:velocity} into Eq.~\eqref{eq:dissipation}, we find that $P=D$, reflecting the absence of dissipation due to quasiparticles in the adiabatic limit. Thus, the current is a pure supercurrent. It transfers 
the energy associated with the voltage 
directly to the spins 
near the domain wall, leading to a residual resistance. This resistance is much lower than the normal-state resistance, because 
there is no Joule heating in the bulk of the superconductor.  
With increasing temperature or for domain wall velocities of the order of $\lambda \Delta/\hbar$, quasiparticles are generated at the domain wall. They provide an additional channel for Gilbert damping \cite{silaev2020finite}.



\section*{Comparison of the normal to the superconducting state}
\label{sec:NtoS}


Besides the superconducting state,  the spin-galvanic effect and its inverse also occur in the normal state. The domain wall can then be driven also above $T_c$.  

As the temperature approaches the critical temperature, the additional Gilbert damping becomes more and more important, leading to significant 
heating within the bilayer. 
This may lead to a local transition to the normal state, over a size 
proportional to the Joule power and inversely proportional to the heat conductance to the substrate and $T\,{-}\,T_c$. When the superconductor 
transitions to the normal state, the domain wall movement becomes much less power-efficient because in addition to the power needed to sustain the motion, power is also required due to the Joule heating in the whole normal part of the setup. In this case, the domain wall movement leads to a small correction of the resistance between two probes (see SI for details),
\begin{align}
    \frac{\Delta R}{R} = 2\frac{\nu_0}{\nu_s}\Big(\frac{G_i l_{\text{so}}}{\sigma}\Big)^2\frac{\hbar\gamma^2  l_{\text{so}}}{C_2 \alpha DL_{\text{probe}}}\;,
\end{align}
where $L_{\text{probe}}$ is the distance between two probes.





\section*{Practical aspects of supercurrent-induced domain wall motion}
\label{sec:outlook}




For both N\'{e}el and Bloch walls, the direction of the domain wall motion is determined by the direction of the current, the sign of the spin-orbit coupling and the direction of magnetization on either side of the wall. 
This has important consequences
for the practical implementation of racetrack memory \cite{parkin2008magnetic}, as adjacent walls move towards or away from each other. 
The mechanism is however useful for controlling the size of different domains in the ferromagnetic insulator \cite{xiang2026supercurrent} and a fast switching of the magnetization in the center of a ferromagnet with two domains, hence providing a memory element.

The strength of the effect depends on the choice of  materials. For a strong coupling between the supercurrent and the domain wall, the superconductor is required to have appreciable spin-orbit coupling. This can be achieved using superconductors with a large spin-Hall angle, such as A15 superconductors \cite{derunova2019giant}, or materials with spin-galvanic effects such as those with gyrotropic crystal structures \cite{he2020magnetoelectric}, or those with Ising spin-orbit coupling. 
The spin-galvanic effect can also be enhanced by adding a non-magnetic heavy atom insulator on top of the superconductor \cite{mazanik2025interfacial}.

Moreover, the ferromagnetic insulator should have a well-defined easy axis, low Gilbert damping, and homogeneous anisotropy, since inhomogeneities lead to domain wall pinning. A good FI/SC interface is also essential. The ferromagnetic insulator should be much thinner than the superconductor. In this case, spin-charge coupling mediated effects naturally dominate over the Oersted field generated by the applied current, with the crossover occurring at around $10$ nm, as discussed in Methods. These constraints make 2D magnets the most promising material candidates. 

For domain wall lengths of order $10$~nm, an equilibrium superfluid momentum of order $\hbar$~\textmu m$^{-1}$, a Gilbert damping parameter $\alpha \sim 10^{-3}$ and thicknesses of order 10 nm, we estimate that the current density required to achieve domain wall speeds of order $100$~m/s, is $10^9$~A/m$^2$, well below the critical current density of many superconductors. The resulting voltage is of order $0.1$~\textmu V. 

The adiabatic limit for the domain wall motion requires $\vdw\ll \lambda \Delta/\hbar$, which for typical setups with $\Delta \sim 0.1$~meV and $\lambda \sim 10$~nm is of order $10^3$~m/s. Accelerating the wall beyond this limit leads to quasiparticle excitation and creation of long-range triplets through time-dependent magnetization \cite{andriyakhina2026long}. They create additional dissipation and thereby decrease the efficiency of domain wall movement. 


Even though in the adiabatic limit there is no dissipation in the superconductor, there is heating in the ferromagnetic insulator due to the Gilbert damping. This heat can either flow into the substrate, or redistribute over the ferromagnetic insulator.
Denoting the interfacial thermal conductivity of the ferromagnetic insulator to the substrate by $\mathcal{K}$ and the thermal conductivity of the FI by $\kappa_{\text{FI}}$, density by $\rho$ and specific heat by $c_p$, the temperature increase in the ferromagnetic insulator is
\begin{align}
    \Delta T &\approx \frac{P}{A_F}\left[\frac{4\kappa\mathcal{K}}{d_{\text{FI}}} + \left(\rho c_p \vdw\right)^2\right]^{-\frac{1}{2}}\;.\label{eq:temperatureincrease}
\end{align}
The temperature increase of the superconductor is of similar order, unless it itself is in better contact with a thermal bath. For typical boundaries, around $1$~K, $\mathcal{K}\sim 10^{2}-10^{4}$~Wm$^{-2}$K$^{-1}$ \cite{swartz1989thermal}, while for EuS $\kappa_{\text{FI}} \sim 10^{-1}$~Wm$^{-1}$K$^{-1}$ \cite{mccollum1964spinwave}, $\rho c_p \sim 10^4$~Jm$^{-3}$K$^{-1}$ \cite{passenheim1966low}.  Using the parameter values above, a domain wall length of the same order as the thickness and a thermal boundary conductance, $\Delta T \sim 1$~mK. Hence, supercurrent control of  domain walls does not lead to major overheating and provides a viable path towards current controlled cryogenic magnetic memory.


\section*{Acknowledgements}

We thank Stuart Parkin and his group members, Vitaly Golovach, and Stefan Ilić for useful discussions. This work was supported by the Research Council of Finland through DYNCOR, Project Number 354735 and through the Finnish Quantum Flagship, Project Number 359240. This work is part of the Finnish Centre of Excellence in Quantum Materials (QMAT). F.~S.~B. thanks financial support from the the Spanish MCIN/AEI/10.13039/501100011033 
through the grant PID2023-148225NB-C31,  the
European Union’s Horizon Europe research and innovation program under grant agreement No.~101130224 (JOSEPHINE), and InstituteQ under the Chair of Excellence programme. 

\bibliography{sources}

\clearpage
\include{methods-NP}

\end{document}

%% file: methods-NP.tex
\setcounter{figure}{0}
\setcounter{equation}{0}
\renewcommand\thefigure{M\arabic{figure}}
\renewcommand\theequation{M\arabic{equation}}

\title{Method section for "Theory of supercurrent induced domain wall movement"}
\maketitle
\section*{Derivation of coupled dynamics from the action}\label{sec:action}
We study a bilayer of a superconductor and a ferromagnetic insulator. We assume that the superconductor can be described within the quasiclassical framework and that it is dirty, that is, that the scattering rate $1/\tau$ is much larger than the superconducting pair potential $\Delta$. In this case, we may describe the superconductor using a non-linear sigma model (NLSM) \cite{efetov1980interaction,wegner1979mobility,finkelshtein1983effect}. Within the NLSM framework, we study a system with a random disorder potential in the Keldysh-Nambu-spin space. 

The random disorder term in the action is decoupled via a Hubbard-Stratonovich transformation. This transformation introduces a $Q$-matrix  \cite{kamenev2023field} and allows for the fermions to be integrated out, so that the action can be considered as a functional of $Q$. In the dirty limit, there is a soft-mode manifold defined by $Q^2 = \mathbf{1}$. This $Q$-matrix at the saddle point corresponds to the quasiclassical Green's function $g$, which means that the saddle point equation of the Keldysh-Nambu-spin NLSM corresponds to the dirty limit kinetic transport equation, that is, the Usadel equation. 

A ferromagnetic insulator can also be described using ann NLSM. In a ferromagnetic insulator, variations in the magnitude of the magnetization are much more expensive than variations in its direction. This defines a soft-mode manifold described by the direction of magnetization $\vec{m}(t,t')$, satisfying $\vec{m}(t,t)^2 = \mathbf{1}$. We define $M(t,t') = \vec{m}(t,t')\cdot\vec{\sigma}$, where $\vec{\sigma}$ is the vector of Pauli matrices in spin space. This way we obtain an NLSM in Keldysh-spin space.

In this manuscript, Pauli matrices in Nambu space are denoted $\tau_{1,2,3}$, while those in Keldysh space are denoted by $\rho_{1,2,3}$. We operate with the Keldysh basis in which all causal configurations are block upper triangular. The quasiclassical Green's function takes the form 
\begin{align}
g(t,t') = \begin{bmatrix}
    g^R(t,t')&g^K(t,t')\\0&g^A(t,t')
\end{bmatrix}\;.
\end{align}
Treating the spins in the ferromagnetic insulator classically, we replace $M(t,t')$ with a time-local field, $M^{\text{cl}}(t)\rho_0 + M^{\text{q}}(t)\rho_1$, where $M^{\text{q}} = 0$ at the saddle point \cite{kamenev2023field}.

The combined action for the electronic and spin NLSMs can be written as:
\begin{align}\label{eq:iS}
   iS[Q,M] = iS_{Q}[Q]+iS_{M}[M]+iS_{I}[Q,M]\;.
\end{align}
For the  superconductor we use the Keldysh NLSM \cite{kamenev2023field,virtanen2021magnetoelectric,virtanen2022nonlinear,kokkeler2025universal}:
\begin{align}
    iS_Q[Q]&= \frac{\pi\nu_0}{2}\text{Tr}\Big[-\frac{D}{4}(\nabla Q)^2+\hat{\omega}_{t,t'}\tau_3Q+\hat{\Delta}Q\\-i\epsilon_{ijl}&\gamma_{lk}\sigma_i Q\sigma_j Q\partial_k Q+\frac{D\theta}{4}\epsilon_{ajk}\sigma_a Q\partial_jQ\partial_kQ \nonumber\\+ \frac{1}{\tau_{\text{so}}}\sigma_k& Q\sigma_k Q\,\Big]\;,\nonumber
\end{align}
where $\text{Tr}$ denotes tracing over matrix degrees of freedom as well as integration over spatiotemporal degrees of freedom, $D$ is the diffusion constant, $\hat\omega_{t,t'} = \delta(t-t')\partial_t$, $\hat{\Delta} = \Delta e^{i\phi\tau_3}\tau_2$ is the pair potential and $\nu_0$ the density of states per spin. We incorporate intrinsic spin-orbit coupling via a spin-galvanic term with spin-galvanic tensor $\gamma_{lk}$, and extrinsic spin-orbit coupling via a spin-Hall term with spin-Hall angle $\theta$. $\tau_{\text{so}}^{-1}$ is the Dyakonov-Perel spin relaxation rate in the normal state, arising from a combination of disorder and spin-orbit coupling. 
For the ferromagnetic insulator we use the action \cite{tatara2008microscopic,hurst2020electron}
\begin{align}\label{eq:actionFI}
    iS_{M} &= \frac{\pi}{2}\nu_s\text{Tr}\,\rho_1\left[-J(\nabla M)^2-K_{\perp} \sigma_z M\sigma_zM\right.\nonumber\\& \left. +K_{\parallel}\sigma_y M\sigma_yM-A[M] \partial_t M+\vec{H}_{\text{ext}}\cdot\vec{\sigma}M\right]\;.
\end{align}
Here $J$ characterizes the exchange stiffness, $K_{\parallel}$ and $K_{\perp}$ denote the easy and hard axis anisotropies respectively, while the last term is a time-derivative term, where $A[M]$ depends on the Berry curvature. The matrix $\rho_1$ appears because the time integral has to be taken over the Keldysh contour, which has opposite orientation on its two branches, see the supplement for more details. 
$\vec{H}_{\text{ext}}$ denotes contributions of an external magnetic field and/or localized spins.
By symmetry, the interactions can be written to second order 
as 
\begin{align}
    iS_{I} = \frac{\pi}{2}\mathrm{Tr}\left(iG_i M\tau_3 Q+ G_r M\tau_3QM\tau_3Q\right)\;.
\end{align}
Here $G_{r,i}$ are the real and imaginary components of the spin-mixing conductance \cite{zhang2019theory}.
The Gilbert damping in the ferromagnetic insulator is usually introduced using a dissipation functional \cite{tatara2008microscopic}:
\begin{align}\label{eq:dissipationfunctional}
    W =  \frac{\pi}{2}\nu_s\text{Tr}\frac{\alpha}{2}(\partial_t M)^2\:.
\end{align}
Together, the actions and dissipation functional in Eqs.~(\ref{eq:iS}--\ref{eq:dissipationfunctional}) describe the coupled dynamics of 
the superconductor and 
ferromagnetic insulator.
\subsection*{Equations of motion}
Because the main contributions to the action stem from configurations near the saddle point, we focus on the saddle point conditions. These yield the equations of motion via the variational derivatives
\begin{align}
    \frac{ \delta S}{ \delta Q} &= 0\;,\\
     \frac{ \delta S}{ \delta M^{\text{q}}} + \frac{ \delta W}{ \delta (\partial_t M^{\text{q}})} &= 0\;,\qquad M^{\text{q}} = 0\;.\label{eq:variationM}
\end{align}
The structural difference between the two conditions arises because the electrons are described using a distribution function, whereas $M$ is a classical, time-local variable.
The normalization conditions restrict the variations to the form $M\xrightarrow{}M+[\beta_M,M]$ and $Q\xrightarrow{}Q+[\beta_Q,Q]$.

The first condition gives the Usadel equation \cite{usadel1970generalized} for the saddle point $g$ of the NLSM, which corresponds to the dirty limit quasiclassical Green's function:
\begin{align}\label{eq:Usadel}
    \nabla \cdot \vec{\mathcal{J}} = [\hat{\omega}_{t,t'}\tau_3 + \hat\Delta,g] +\mathcal{T}_1\; ,
\end{align}
where $\vec{\mathcal{J}}$ is the matrix current given by
\begin{align}\label{eq:current}
    \mathcal{J}_k &= -Dg\partial_k g +\frac{i}{16}\epsilon_{ijl}\gamma_{lk}\{[\sigma_i,g],\sigma_j+g\sigma_j g\}\nonumber\\&+\frac{D\theta}{4}\epsilon_{ajk}\{\partial_jg,\sigma_a + g\sigma_a g\}\;,
\end{align}
while $\mathcal{T}_1$ is the matrix torque due to spin-orbit coupling
\begin{align}
    \mathcal{T}_1 &= -\frac{i}{8}\epsilon_{ijl}\gamma_{lk}[\{\partial_k g,g\sigma_i g\},\sigma_j] \nonumber\\&-\frac{D\theta}{4}\epsilon_{ajk}[\sigma_a,Q\partial_j Q\partial_k Q]+\frac{1}{\tau_{\text{so}}}[\sigma_k Q \sigma_k, Q]\;.
\end{align}
Additionally, the variation directly gives the boundary condition
\begin{align}\label{eq:BC_Usadel}
    \vec{n}\cdot\vec{\mathcal{J}} = \mathcal{T}_2\;,\;
\end{align}
where $\vec{n}$ is the normal vector at the interface pointing outward from the superconductor, and the torque $\mathcal{T}_2$ is given by
\begin{align}
\mathcal{T}_2 &= iG_i [M\tau_3, g]+ 2G_r [M\tau_3gM\tau_3,g]\;.
\end{align}
The saddle point equation for $M$ gives the Landau-Lifshitz-Gilbert equation,
\begin{align}
    \partial_t M + \alpha [M,\partial_t M]&= J\nabla (  M\nabla M) + [K_\parallel\sigma_yM\sigma_y,M]\nonumber\\-[K_{\perp}\sigma_z &M\sigma_z,M]+[\vec{H}_{\text{ext}}\cdot\vec{\sigma},M]\;,
\end{align}
with the boundary condition
\begin{align}
    \vec{n}\cdot(-JM\nabla M)= -\mathcal{T}_S\;,
\end{align}
in which the interaction terms contribute a torque
\begin{align}\label{eq:Newton}
    \mathcal{T}_{S} = \int_{-\infty}^{\infty} d(t-t')\text{tr}_{\rho,\tau}\rho_1 \mathcal{T}_2\;.
\end{align}
The notation $\text{tr}_{\rho,\tau}$ indicates that traces are only taken over the Keldysh and Nambu degrees of freedom, so that the spin degree of freedom remains. Equations (\ref{eq:BC_Usadel}) and (\ref{eq:Newton}) reflect Newton's action--reaction principle, showing that the torque exerted on the conduction electrons of the superconductor is equal and opposite to the torque experienced by the spins in the ferromagnetic insulator.

\section*{Equations for the collective domain wall coordinates}\label{sec:Domainwallcoordinates}

In this section, we discuss how the combined LLG and Usadel equations can be transformed into equations of motion for the effective domain wall coordinates.

When $K_{\parallel}\gg K_{\perp}$, the domain wall shape is rigid, and thus the wall can be parametrized using effective collective coordinates \cite{tatara2008microscopic,hurst2020electron}: the position $x_0$ of the domain wall, 
and the angle $\phi$ of magnetization rotation around the $x$ axis. We can restrict our analysis to perturbations that either translate the domain wall position, 
\begin{align}
\vec{m}\xrightarrow{}\vec{m}+\rho_1(\partial_x \vec{m}) \delta x_0\;,
\end{align}
or rotate the spins around the easy axis (here the $y$-axis). The latter affects only the orientation of spins inside the wall:
\begin{align}
\vec{m}\xrightarrow{}\vec{m}+\rho_1(\vec{\hat{y}}\times \vec{m})\delta \phi\;.
\end{align}
These correspond to variations 
$\beta_M = i\rho_1 [M,\partial_x M]\delta x_0$ and $\beta_M = i \rho_1 [M,[\sigma_y,M]]\delta \phi$, respectively. The first Pauli matrix $\rho_1$ is required in both because the variation of magnetization is taken with respect to the $\rho_1$ component in Eq.~(\ref{eq:variationM}).

The variation of the time-derivative, spatial derivative and anisotropy terms give the standard domain wall equations while variation of $W$ with respect to $\partial_t \phi$ and $\partial_t x_0$ gives the usual Gilbert damping terms. Explicit calculations for these terms are given in the Supplemental Material, here we focus on the terms that arise from the boundary action.

The variation of the interaction term, $i\delta S_{I}= \frac{\pi}{2}(F \delta x_0 + \Tau\delta\phi)$, gives the domain wall force and torque:
\begin{align}
    F[g] &=  \int_{-\infty}^{\infty} dx\,\text{tr}\rho_1 \mathcal{T}_2[g] [M,\partial_x M]\;,\\
    \Tau[g] &= \int_{-\infty}^{\infty} dx\,\text{tr}\rho_1 \mathcal{T}_2[g] [M,[M,\sigma_y]]\;.
\end{align}
Here $F$ has units of force per unit area, while $\Tau$ has units of torque per unit area. As an example, evaluating the $G_i$ term gives Eqs. \eqref{eq:tau} and \eqref{eq:force} in the main text:
\begin{align}
F &= \frac{4G_i}{e^2\nu_0} \int_{-\infty}^{\infty} dx\, \vec{S}(x) \cdot \partial_x \vec{m}\;,\\
\Tau &= \frac{4G_i}{e^2\nu_0} \int_{-\infty}^{\infty} dx \,\vec{S}(x)\cdot [ \vec{m}\times \hat{\vec{y}} ]\;,
\end{align} 
where 
$\vec{S}(x) = \nu_0\int_{-\infty}^{\infty} dE\,\text{tr}_{\rho,\tau}\tau_3 g(x)$
is the excess electron spin density 

Combining the bulk and interfacial variations results in two coupled equations of motion for the domain wall coordinates: 
\begin{align}\label{eq:CC1}
   m_y^R\nu_s \partial_t x_{0}- C_1\nu_s\alpha \lambda\partial_t \phi \hspace{3em}&\\ + C_1\nu_s\lambda K_{\perp} \sin 2\phi  +  \frac{\Tau[g]}{d_{\mathrm{FI}}}&= 0\;,\nonumber\\ 
   m_y^R \nu_s \partial_t \phi + C_2\nu_s \frac{\alpha}{\lambda} \partial_t x_{0}+ \frac{F[g]}{d_{\mathrm{FI}}}&= 0\;.\label{eq:CC2}
\end{align}
These equations are exactly of the same form as the 
equations in the ballistic limit \cite{tatara2004theory}, the only difference is that the input forces now depend on the $g$ matrix. 
Solving the domain wall coordinate equations, Eqs.~(\ref{eq:CC1}--\ref{eq:CC2}), in combination with the Usadel equation, Eq.~(\ref{eq:Usadel}), gives the full domain wall dynamics. By dividing Eq.~(\ref{eq:CC1}) by $\lambda\nu_s$ and Eq.~(\ref{eq:CC2}) by $\nu_s$ one obtains Eqs. \eqref{eq:DWeq1} and \eqref{eq:DWEq2} in the main text. 

\section*{Solution of the equations of motion}\label{sec:Solutiondomainwalldynamics}
In this section, we solve the domain wall equations and relate the domain wall velocity to the force exerted by the electrons. 
\subsection*{Domain Wall Force effects}
In the configuration studied in the main text, the torque $\Tau = 0$. 
Equations (\ref{eq:CC1}--\ref{eq:CC2}) can be reduced to a single equation by solving for $\partial_t x_0$ in the second equation and substituting it into the first, yielding
\begin{align}
    (1+C_1C_2\alpha^2) \partial_t \phi + C_1 C_2\alpha K_\perp \sin 2\phi = -\frac{m_y^R F}{\nu_s \dFI}\;.
\end{align}
This equation is analogous to the Resistively Shunted Junction (RSJ) model of Josephson junctions \cite{hurst2020electron}, with the angle $\phi$ playing the role of the phase difference.


Below the Walker breakdown threshold current playing the role of the critical current, $|F| < F_t = \nu_s\dFI C_1\alpha K_\perp$, $\partial_t \phi = 0$ in the steady state. However, $\phi$ and $\langle \partial_t x_0\rangle$ do not vanish. Instead, here the domain wall tilts at an angle and moves at a steady speed without precession:
\begin{align}
    \sin 2\phi_0 = -m_y^R\frac{F}{F_t}\;,\qquad
    \partial_t x_0 = -\frac{\lambda}{C_2\alpha}\frac{F}{\nu_s \dFI}\;.\label{eq:vdwsmallI}
\end{align}

When $|F| > F_t$, a precessional motion sets in:
\begin{align}
    \langle \partial_t \phi\rangle = -m_y^R F\frac{\sqrt{ 1- (F_t/F)^2}}{\nu_s \dFI(1+C_1C_2\alpha^2)}\;.
\end{align}
In this regime, the average domain wall speed is reduced:
\begin{align}
   \vdw  &= \langle \partial_t x_0\rangle = - \frac{\lambda}{C_2\alpha}\left(\frac{F}{\nu_s \dFI} + m_y^R \langle\partial_t \phi\rangle\right). 
\end{align}
Asymptotically, when $F\gg C_1\alpha \nu_s \dFI K_{\perp}$, 
the domain wall motion becomes steady, with constant velocity and precession frequency. 

\section*{Domain wall force and torque balance in materials with a spin-Hall effect}\label{sec:torqueandbalanceHall}
In this section, we derive the domain wall force that arises in the case of a superconductor with a spin-Hall angle $\theta$. Specifically, we calculate the spin density generated through the spin-Hall effect and the resulting torque between the electrons in the superconductor and the spins in the ferromagnetic insulator. We show that the torque of the spins on the electrons is exactly counterbalanced by a generalized spin-orbit coupling torque, consistent with the absence of spin currents on either side of the domain wall. Specifically,  
in the normal state the torque on the electrons is counterbalanced by the spin-relaxation torque, while in the superconducting state it is counterbalanced by the spin-Hall torque. For simplicity of presentation, we do not include intrinsic spin-orbit coupling and focus on the superconducting state. Calculations for the normal state and intrinsic spin-orbit coupling are discussed in the Supplementary Material.

To illustrate the balance of the torques below $T_c$ we first consider the limit close to $T_c$. Afterwards, we discuss the low temperature limit. Near $T_c$, the Usadel equation can be linearized in the order parameter $\Delta$, with the quasiclassical Green's function approximated as $g \approx \tau_3 + f\tau_+ + \tilde f\tau_-$, where $f$ and $\tilde f$ are the anomalous parts of the Green's function, and the ladder operators are defined as $\tau_\pm = \frac 1 2 (\tau_1 \pm i \tau_2)$. In the adiabatic limit, the torque can be calculated using equilibrium Green's functions, which we treat using the Matsubara technique.

We consider the solution to the Usadel equation, in a bulk superconductor with an interface at $z  = 0$ with vacuum and at $z = \dSC$ with a homogeneous $\hat m$ in the adjacent FI, to linear order in $q_{\mathrm{ex}}$ and the first order in $G_i$. In this case, the anomalous functions are $f = (f_s + \vec{f}_t\cdot\vec{\sigma})e^{iq_{\text{ex}}x}$, $\tilde{f} = (f_s + \vec{f}_t\cdot\vec{\sigma})e^{-iq_{\text{ex}}x}$, where $f_s = \Delta/\omega$ describes the singlet correlations, and the triplet correlations are given by
\begin{align}
    \vec{f}_t =  \frac{G_i\xi_t}{\sigma}\frac{\Delta}{\omega}\frac{\cosh{\frac{z}{\xi_t}}}{\sinh{\frac{\dSC}{\xi_t}}}\vec{m} + \theta q_{\text{ex}} \xi_t \frac{\sinh{\frac{z-\frac{\dSC}{2}}{\xi_t}}}{\cosh{\frac{\dSC}{2\xi_t}}}\frac{\Delta}{\omega}\vec{\hat{y}}\;,
\end{align}
where $\omega$ is the Matsubara frequency, $d$ is the thickness of the superconductor and $\xi_t$ is the the coherence length of the triplet pairs. 
From this solution, the torque between the electrons in the superconductor and spins in the ferromagnetic insulator at the SC-FI interface located at $z=\dSC$ can be calculated up to second order in $\Delta$
\begin{align}
    \mathcal{T} &= 
    G_i \vec{m}\times\frac{\pi T}{4}\sum_\omega\text{tr}[\vec{\sigma}\tau_3 g(\omega; z=\dSC)]\nonumber\\
    &\approx \theta G_i q_{\text{ex}} (\vec{m}\times \hat{\vec{y}}) \pi T\sum_\omega \xi_t\frac{\Delta^2}{\omega^2}\tanh{\frac{\dSC}{2\xi_t}}\;.
\end{align}
The spin-Hall torque on the other hand reads
\begin{align}\label{eq:spinhall_torque}
   \mathcal{T}_\theta  &= \sigma\frac{\theta}{4}\sum_\omega\text{tr}[\sigma_yQ(\partial_x Q\partial_z Q-\partial_z Q\partial_x Q)]\\
   &\approx
   \sigma\theta q_{\mathrm{ex}} \vec{\hat{y}}\times\partial_z (f_s \vec{f}_t)\;. 
\end{align}
By integrating over the thickness, we find that
\begin{align}
\int_{0}^{d_{\text{SC}}} dz \mathcal{T}_\theta = -\mathcal{T}\,,
\end{align}
demonstrating that the spin-Hall torque exactly compensates the torque from the coupling to the ferromagnetic insulator. Consistent with this, we find that in equilibrium 
\begin{align}
    \mathcal{T}_{\text{SR}} = \tau_{\text{so}}^{-1}\text{tr}\sigma_a [\sigma_k g\sigma_k ,g] = 0\;.
\end{align}
meaning that there is no torque from the spin-orbit relaxation. 

Thus, the torque balance in the superconducting state is completely different compared to the normal state, in which spin-Hall torques of Eq.~\eqref{eq:spinhall_torque} vanish and the spin-relaxation torque absorbs the torque of the ferromagnetic insulator.

Next, we consider the same type of calculation, but for temperatures  well below the critical temperature. In this case we may expand $g\approx g_0 + \delta g$, with $g_0$ being the zero current quasiclassical Green's function, while $\delta g$ now satisfies the differential equation
\begin{align}
    \partial_{z}^2\delta g &= \delta g/\xi_t^2\;,\\
    \partial_{z}\delta g &=\theta q_{\text{ex}} g_0 [\tau_3,g_0]\;,  & z &= d_{\text{SC}}\;,\\
    \partial_{z}\delta g &=\theta q_{\text{ex}} g_0 [\tau_3,g_0]\;,& z &= 0\;,
\end{align}
where $\xi_t^2 = D/[2(\sqrt{\omega^2 + \Delta^2}+\tau_{\text{so}}^{-1})]$. The solution is
\begin{align}
    \delta g = \theta q_{\text{ex}} \xi_t \frac{\Delta^2}{\omega^2+\Delta^2} \frac{\sinh{\frac{z-\frac{\dSC}{2}}{\xi_t}}}{\cosh{\frac{\dSC}{2\xi_t}}}\;.
\end{align}
Thus, the spin at the interface is
\begin{align}
    S_y &= 
    \nu_0\theta q_{\text{ex}} \pi T\sum_\omega \xi_t\tanh{\frac{\dSC}{2\xi_t}}\frac{\Delta^2}{\omega^2+\Delta^2}\;. 
\end{align}
If the spin-orbit relaxation length is shorter than the singlet superconductor coherence length,
\begin{align}
    S_y&= \hbar\nu_0\theta q_{\text{ex}} l_{\text{so}}\tanh{\frac{\dSC}{2l_{\text{so}}}} \pi T\sum_\omega \frac{\Delta^2}{\omega^2+\Delta^2}\nonumber\\ &= \nu_0 \kappa \dSC \pi T \sum_\omega \frac{\Delta^2}{\omega^2+\Delta^2}\;,
\end{align}
independent of the relative scales of $\Delta, T, \tau_{\text{so}}^{-1}$.
Using $j = D_S q_{\text{ex}} = \sigma\pi T q_{\text{ex}}\sum \omega \frac{\Delta^2}{\omega^2 + \Delta^2}$ this can be written as
\begin{align}
    S_y =  \chi_s \kappa j\;,
\end{align}
where $\chi_s = \frac{\hbar}{e}\frac{ d_{\text{SC}}}{D}$ is a local spin response coefficient.

\section*{Equilibrium superfluid momentum}\label{sec:q0}
In this section we explicitly calculate the equilibrium phase gradients in the superconductor. To this end, we first calculate the anomalous current, and then calculate the phase gradient needed to cancel this anomalous current using the usual contribution.

Since the anomalous currents are already linear in the spin-orbit coupling coefficient, only the zero-order Green's functions are needed at this order, and these coincide for the inverse spin-galvanic and spin Hall situations. They satisfy
\begin{align}
    D \partial_z (g\partial_{z} g) &= [\omega \tau_3 + \hat{\Delta} + \tfrac{1}{2}\tau_{\text{so}}^{-1}\sigma_k g\sigma_k, g]\;,\mspace{-40mu}\\
    \sigma g\partial_z g & = G_i[\vec{m}\cdot\vec{\sigma}\tau_3,g]\;, & z &= d_{\text{SC}}\;,\\
    \sigma g\partial_z g & = 0\;,& z &= 0\;.
\end{align}
To first order in the spin-mixing conductance $G_i$, the solution to this equation reads 
\begin{align}
    g &= g_0 \sigma_0 + g_1 \vec{m}\cdot\vec{\sigma}\;,\\
    g_0 &= \frac{\omega\tau_3 + \Delta\tau_2}{\sqrt{\omega^2+\Delta^2}}\;,\\
    g_1 &=\frac{2G_i \xi_t}{\sigma}\frac{\Delta(\omega\tau_2 -\Delta \tau_3)}{\omega^2+\Delta^2}\frac{\cosh\frac{z}{\xi_t}}{\sinh{\frac{d_{\text{SC}}}{\xi_t}}}\;.
\end{align}
Inserting this into Eq.~\eqref{eq:current} provides the anomalous current.

\subsection*{Intrinsic SOC}
In the case of intrinsic SOC, the anomalous current is given by 
\cite{kokkeler2024nonreciprocal}
\begin{align}
    j_{\text{an}} &= \nu_0 \gamma\frac{1}{d_{\text{SC}}}\int_{0}^{d_{\text{SC}}}dz \sum_\omega  \frac{1}{2}\text{tr}\tau_3\sigma_y g \nonumber\\
    &\approx \nu_0\gamma\frac{G_il_{\text{so}}^2}{d_{\text{SC}}\sigma}m_y  \sum_\omega \frac{\Delta^2}{\omega^2+\Delta^2}\;,
\end{align}
where we assumed a large spin-orbit relaxation rate, so that $\xi_t \approx l_{\text{so}}$, with $l_{\text{so}} = \sqrt{D\tau_{\text{so}}}$.
Since the usual supercurrent is given by $j_s = -\nu_0 D q \sum_\omega \frac{\Delta^2}{\omega^2+\Delta^2}$, the total current vanishes at equilibrium superfluid momentum $\hbar q=\hbar q_0$ with
\begin{align}
    q_0 \approx \frac{\gamma}{D}\frac{G_il_{\text{so}}^2}{d_{\text{SC}}\sigma}m_y \approx \gamma \tau_{\text{so}}\frac{G_i}{d_{\text{SC}}\sigma}m_y\;.
\end{align}

\subsection*{Spin-Hall angle}
Similarly, in the case of a spin-Hall angle, 
in the limit where the spin-orbit relaxation length is much smaller than the singlet superconductor coherence length, the equilibrium superfluid momentum is given by \cite{virtanen2021magnetoelectric}
\begin{align}
    q_0 = \theta \frac{G_i}{\sigma}\frac{l_{\text{so}}}{d_{\text{SC}}}\tanh\frac{d_{\text{SC}}}{2l_{\text{so}}}\;,
\end{align}
which holds as long as the thickness is much shorter than the superconductor coherence length. 

\section*{Voltage}\label{sec:voltage}
In this section we relate the equilibrium phase gradient to a voltage. To this end, we need to find out the chemical potential difference between the two sides of the superconductor.

In what follows, we choose a gauge in which $\partial\vec{A}/\partial t = 0$. In this case, the measured voltage can be obtained purely from the chemical potential. The potential is defined as $\mu(r) = \int_{-\infty}^{\infty} dE \text{tr}g^K(r,E)$, where $r$ is a coordinate in the normal wire. This potential however is gauge dependent, the physically relevant quantity to consider is $\mu_N = \mu + \Phi$, where $\Phi$ is the scalar potential. We exploit this gauge dependence by using space independent gauge transformations, which do not affect $\vec{A}$, but transform the pair potential as $\Delta \xrightarrow{}\Delta e^{i\chi(t)}$ and the scalar potential as $\Phi \xrightarrow{}\Phi + \partial_t \chi$.

We may therefore find a gauge $\Phi_{R}$ in which the pair potential is time independent. In this gauge, the superconductor to the right of the domain wall is a superconductor with time-independent phase, and therefore within this gauge $\mu_N(r\xrightarrow{}\infty, \Phi = \Phi_R) = 0$. Similarly, there exists a gauge $\Phi_L$ so that the pair potential on the left is time-independent, and by similarity we find $\mu_N(r\xrightarrow{}-\infty, \Phi = \Phi_L) = 0$. Therefore, we find \begin{align}\mu_N(r_R,\Phi) -\mu_N (r_L,\Phi) = \Phi_R-\Phi_L\;.\end{align} 

We thus need to determine $\Phi_R-\Phi_L$ by eliminating the time-dependence of the pair potential, i.e.,
\begin{align}
    \Phi_L-\Phi_R = i \left[\partial_t \Delta (x\xrightarrow{}-\infty)-\partial_t \Delta (x\xrightarrow{}\infty)\right]\;.
\end{align}
To determine their values, we consider the time-dependent Usadel equation:
\begin{widetext}
\begin{align}
    \{\tau_3,\partial_t g\} &= [-D(\nabla^2 g) + E\tau_3+\check{\Delta} + \vec{h}(x-\vdw t)\cdot\vec{\sigma}\tau_3 + \Gamma \tau_3\vec{\hat{h}}(x-\vdw t)\cdot\vec{\sigma}\circ g\circ \tau_3\vec{\hat{h}}(x-\vdw t)\cdot\vec{\sigma}\commcirc g]\nonumber\\&+ \frac{i\gamma}{8}\partial_x \epsilon_{ijy} \Big(\{[\sigma_i, g]\commcirc \sigma_j + g \sigma_j \circ g\}\Big)+\frac{i\gamma}{16}\epsilon_{ijy}[\{\partial_x g\commcirc g\sigma_j \circ g\},\sigma_j] + \mathcal{I}\;.
\end{align}
Here the symbol $\circ$ denotes a convolution in time-domain and $ \mathcal{I}$ represents the collision integral for inelastic processes. We do not consider its explicit form here.

Let us consider a transformation to a co-moving frame, by replacing $x\xrightarrow{}\Tilde{x}\equiv x + \vdw t$. In that case, by Galilean invariance there exists a gauge so that the solution does not depend explicitly on $t$ anymore, and we have
\begin{align}
    -\vdw \{\tau_3,\partial_{\Tilde{x}} g\} &= [-D(\tilde{\nabla}^2 g) + E\tau_3+\check{\Delta} + \vec{h}(\tilde{x})\cdot\vec{\sigma}\tau_3+ \Gamma \tau_3\vec{\hat{h}}(\tilde{x})\cdot\vec{\sigma}\tilde{\circ} g\tilde{\circ} \tau_3\vec{\hat{h}}(\tilde{x})\cdot\vec{\sigma}\tilde{\circ} g\tilde{\commcirc} g]\nonumber\\&+ \frac{i\gamma}{8}\epsilon_{ijy} \partial_x \Big(\{[\sigma_i, g]\tilde{\commcirc} \sigma_j + g \sigma_j \tilde{\circ} g\}\Big)+\frac{i\gamma}{16}\epsilon_{ijy}[\{\partial_x g\tilde{\commcirc} g\sigma_j \tilde{\circ} g\},\sigma_j] + \tilde{\mathcal{I}}\;.
\end{align}
\end{widetext}
Here, $\tilde{\circ}$ is the convolution written in the new coordinates. Its Moyal product has $\vdw \partial_{\tilde{x}}$ instead of $\partial_t$, as can be verified through the chain rule. Similarly $\tilde{\mathcal{I}}$ corresponds to $\mathcal{I}$ with the appropriate modification of the convolutions through the chain rule.
Moreover, using the chain rule, we find
\begin{align}
    \Phi_R-\Phi_L &= \\
    \vdw \text{Im}&\Delta^{-1}\left[ \partial_{\tilde{x}} \Delta (x\xrightarrow{}-\infty)-\partial_{\tilde{x}} \Delta (x\xrightarrow{}\infty)\right]\;.\nonumber
\end{align}

In the adiabatic limit, $\vdw\ll \lambda \Delta/\hbar$, we may evaluate $\partial_{\tilde{x}}$ for the zeroth order solution, which gives $\text{Im}\Delta^{-1}\partial_{\tilde{x}}\Delta(\xrightarrow{}\pm \infty) = \pm q_0$, where $q_0$ is the equilibrium phase gradient. With this, we find the potential drop of the superconducting condensate to be
\begin{align}\label{eq:PhiLPhiR}
    \Phi_L-\Phi_R = 2\vdw q_0\;.
\end{align}
A voltage probe, however, measures the potential of the quasiparticles. Close to the domain wall, these two potentials can be different due to charge imbalance between the quasiparticles and the superconducting condensate \cite{tinkham1996introduction,heikkila2019thermal,tokatly2025spin}. The charge imbalance mode has short-ranged ($\sim \lambda,\xi$) contributions below the gap and long-ranged (inelastic scattering length) contributions above the gap. However, for low temperatures, the latter type of contributions are suppressed as $e^{-\Delta/T}$. Therefore, if the probes are far away from the domain wall on the scale of the superconductor coherence length, the measured voltage is
\begin{align}
    V = \frac{\hbar}{2e}(\Phi_L-\Phi_R) = \frac{\hbar}{e}\vdw q_0\;.
\end{align} The charge imbalance does lead to a tail in the measured voltage after the domain wall has left the space between the two probes.

\section*{Power and dissipation}\label{sec:PowerandDissipation}
In this Section we calculate the power required to sustain the charge current and the dissipation due to Gilbert damping and show that they are equal. With this we derive Eqs. \eqref{eq:Power} and \eqref{eq:dissipation} in the main text
\subsection*{Power}
First we consider the power required to sustain the charge current. From the solution to the Usadel equation, one can compute the gauge invariant charge ($j_c$) and heat ($j_Q$) currents
\begin{align}
    j_c &=  \text{tr}\rho_1\tau_3 (g\nabla g)(t,t)\;,\\
    j_Q &= \Phi j_c + j_E\;,\\
    j_E & = \text{tr}\rho_1(\partial_{t_1}-\partial_{t_2}) (g\nabla g)(t_1,t_2)|_{t_1 = t_2}\;,
\end{align}
where $\Phi$ is the scalar potential and $j_E$ the energy current.
The power delivered by the electrons is the difference in the heat carried by them on the left and right side of the domain wall,
\begin{align}
    P = j_Q(x\xrightarrow{}-\infty)-j_Q(x\xrightarrow{}\infty)\;.
\end{align}

Now consider the domain wall system. We start from a gauge in which $\partial \vec{A}/\partial t = 0$. On the left, far away from the domain wall (much further than the inelastic scattering rate), the solution converges to that of a bulk superconductor with a supercurrent $j_c$ and energy current $j_{E0}$. This means that we can do a uniform gauge transformation to a gauge in which $\vec{A} = 0$, $\Delta$ is time-independent and $g$ only depends on the time-difference coordinate on the far left side. We denote the scalar potential in this gauge by $\Phi = \Phi_{L}$. 

Similarly, on the right hand side, far away from the domain wall, we can do a uniform gauge transformation $\vec{A} = 0$ and in which $\Delta$ is time-independent and $g$ only depends on the time-difference on the right. We denote the scalar potential in this case by $\Phi = \Phi_R$.  Since the scalar potential does not enter the Usadel equation, the problem is entirely similar to the one for the left, so that in this gauge the energy current on the right side is $j_{E0}$.

From this we conclude that
\begin{align}
    \frac{2e^2}{\hbar}P &= j_{Q}(x\xrightarrow{}\infty)-j_{Q}(\xrightarrow{}\infty) \\&= j_{E0}+\Phi_L j_c- (j_{E0}+ \Phi_R j_c) = (\Phi_L-\Phi_R)j_c\;.\nonumber
\end{align}
Using Eq.~\eqref{eq:PhiLPhiR}, we find that
\begin{align}\label{eq:P}
    P = 2A\frac{\hbar}{e}\vdw q_0 j\ = A\frac{\hbar}{e^2}\frac{q_{0}^2\lambda}{C_2\alpha}\frac{d_{\text{SC}}}{d_{\text{FI}}}\frac{j^2}{e \nu_s}\;,
\end{align}
confirming  Eq.~(\ref{eq:Power}) in the main text.

\subsection*{Dissipation and balance to power}
Next, we consider the dissipation due to Gilbert damping and show that it equals the power required to keep the current flowing. We consider the case in which the current is below the Walker breakdown current. In this case only $\partial_t x_0$ is nonzero. We have
\begin{align}
    D &= \hbar \nu_s \AF \alpha \int_{-\infty}^{\infty} dx (\partial_t \vec{m}) = \AF C_2\hbar \alpha\frac{\nu_s}{\lambda} \vdw^2\nonumber\\&= A \hbar\frac{q_{0}^2\lambda}{C_2\alpha}\frac{d_{\text{SC}}}{d_{\text{FI}}}\frac{j^2}{e^2 \nu_s} = P\;,
\end{align}
with $A/\AF = \dSC/\dFI$. This is Eq. \eqref{eq:dissipation} in the main text and shows the balance between power and dissipation below the breakdown. The equality also holds above it.

\section*{Motion induced by the Oersted field}\label{sec:Oersted}

\begin{figure}[h]
    \centering
    \includegraphics[width=\linewidth]{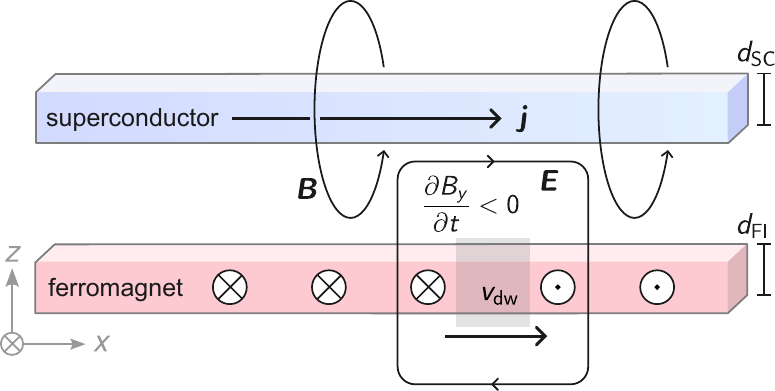}
    \caption{Electrodynamically driven domain wall motion in parallel SC and FI wires. The Oersted field of the supercurrent exerts a torque on the FI, driving the domain wall. A moving domain wall is accompanied by a time-dependent magnetic field, which induces a solenoidal electric field.}
    \label{fig:electrodynamic}
\end{figure}

Even in the absence of spin-charge coupling, there is a mechanism to move domain walls with supercurrents. Following Maxwell's equations, the flow of a supercurrent creates an Oersted field encircling the current flow, as illustrated in Fig. \ref{fig:electrodynamic}. At the position of the ferromagnetic insulator, this Oersted field points along the $\pm \vec{\hat{y}}$ direction, which is exactly the right direction for domain wall movement in our configuration. Since magnetic fields $\vec{H}$ enter the LLG equation as $\mu_B \vec{m}\times \vec{H}$, this Oersted field makes the domain wall move in the same way as the spin-orbit coupling induced spin. Moreover, through the Faraday effect, the domain wall movement creates an electric field that needs to be overcome by the power source. Thus, also Oersted induced motion leads to a finite resistance, with a voltage drop near the domain wall. This makes the effect very similar to the spin-orbit coupling induced effect.

Which of the two effects dominates depends on the geometry of the setup. The Oersted field that is produced by the current has magnitude
\begin{align}
    B_\phi = \frac{\mu_0 I}{2\pi r} = \frac{\mu_0 d_{SC}W_{SC}}{2\pi r}j\;,
\end{align}
where $W_{SC}$ is the width of the superconductor and $r$ is the distance to its core. In the LLG action, the quantity $\nu_s \vec{h}\cdot\vec{m}$ appears, where $\vec{h} = g\mu_B B$ is the effective exchange field and $\nu_s$ is the density of spins.

Integration over the volume now gives the following torque per unit width of the FI:
\begin{align}
    \mathcal{T}_1 & = \frac{e}{\hbar}\frac{\nu_s g\mu_B\mu_0 d_{SC}W_{SC}}{2\pi} \text{ln}\Big(1+\frac{2d_{FI}}{d_{SC}}\Big)\Big(\frac{\hbar}{e} j\Big)\;,
\end{align}
where the final expression was written such that the prefactor is dimensionless.
This has to be compared by the torque induced by spin-orbit coupling
\begin{align}
    \mathcal{T}_2 = \frac{G_i l_{\text{so}}}{\sigma} \Big(\frac{\hbar}{e} j\Big)\;.
\end{align}
The spin-orbit mechanism dominates if $\mathcal{T}_2>\mathcal{T}_1$. This directly gives the following cross-over condition

\begin{align}
    \theta\frac{G_i l_{\text{so}}}{\sigma}\frac{2\pi\hbar}{e\mu_0 M A\,\text{ln}(1 + \frac{d_{\text{FI}}}{d_{\text{SC}}})}>1\;,\label{eq:crossovercondition}
\end{align}
where $M = \nu_s g\mu_B$ is the magnetization density of the ferromagnetic insulator. Taking typical values of $G_i \sim 10^{13}$~S/m${}^2$, $\sigma \sim 10^7$~S/m, the cross-over between Oersted dominated and spin-orbit dominated movement happens when the spatial dimensions are of order 10~nm. 

In practice both mechanisms are present simultaneously. Unlike the rapidly decaying spin-orbit coupling induced motion, the Oersted field decays slowly and can lead to cross-talk in systems with multiple S/FI systems.

\section*{Temperature increase}\label{sec:temperature}
The dissipation due to Gilbert damping in the ferromagnetic insulator leads to heating of the ferromagnetic insulator and the superconductor close to the domain wall.
In this Section we calculate the temperature increase in the ferromagnetic insulator due to the dissipation. 
In the calculation, we ignore the heat transfer to the superconductor for simplicity of presentation. Moreover, since thermal length scales are typically much larger than the domain wall length (see SI), we approximate the domain wall as a $\delta$ function source. For a steady movement, in the co-moving frame the temperature difference with the substrate $\Delta T(x)$ obeys 
\begin{align}
    \kappa \partial_{x}^2 \Delta T  - \rho c_p \vdw \partial_x T  = -\frac{D}{A_F} \delta(x) + \frac{\mathcal{K}}{ d_{\mathrm{FI}}} \Delta T\;.\label{eq:thermal1Ddelta}
\end{align}
The first term on the left represents the usual diffusion term, while the second term stems from the time-derivative term. On the right side, we have included the dissipation term as well as the transfer of heat to the substrate.
This equation is readily solved,
\begin{align}
    \Delta T(x) = 
    \begin{cases}
    \Delta T(0) e^{\frac{x}{l_{T+}}}\;,&x<0\;,\\
    \Delta T(0) e^{\frac{-x}{l_{T-}}}\;,&x>0\;.
    \end{cases}
\end{align}
where the effective decay lengths are $l_{T\pm}^{-1}= \sqrt{\left(\frac{\rho c_p \vdw }{2\kappa}\right)^2 + \frac{\mathcal{K}}{\kappa d_{\mathrm{FI}}}}\pm \frac{\rho c_p\vdw }{2\kappa}$, and 
the temperature increase at the domain wall is
\begin{align}
    \Delta T(0) 
    &= 
    \frac{D}{A_F}\left[\frac{4\kappa\mathcal{K}}{d_{\text{FI}}} + \left(\rho c_p \vdw\right)^2\right]^{-\frac{1}{2}}\;.
\end{align}
Here we also used $D = P$. This is Eq.~\eqref{eq:temperatureincrease} in the main text. This also provides an upper limit to the temperature increase in the superconductor.